\documentclass[conference]{IEEEtran}
\IEEEoverridecommandlockouts
\usepackage{cite}
\usepackage{amsmath,amssymb,amsfonts}
\usepackage{algorithmic}
\usepackage{graphicx}
\usepackage{textcomp}
\usepackage{xcolor}
\usepackage{makecell}

\usepackage{siunitx}
\usepackage{hyperref}
\usepackage[nolist]{acronym}
\usepackage[flushleft]{threeparttable}
\usepackage{placeins}
\usepackage{svg}                                   
\usepackage[percent]{overpic}

\usepackage{mathtools}
\usepackage[export]{adjustbox}
\usepackage{booktabs}

\usepackage{eso-pic}
\usepackage{url}

\def\BibTeX{{\rm B\kern-.05em{\sc i\kern-.025em b}\kern-.08em
    T\kern-.1667em\lower.7ex\hbox{E}\kern-.125emX}}
\begin{document}
\bstctlcite{IEEEexample:BSTcontrol}

\newcommand{\todo}[1]{\textcolor{red}{#1}}                          
\newcommand{\done}[1]{\textcolor{Green}{#1}}                        
\DeclareSIUnit{\dBm}{\text{dBm}}                                    
\newcommand{\RNum}[1]{\uppercase\expandafter{\romannumeral #1\relax}}  

\newcolumntype{P}[1]{>{\centering\arraybackslash}p{#1}}
\newcommand{\bfqty}[2]{\text{\bfseries\SI{#1}{#2}}}
\renewcommand*{\figureautorefname}{Fig.}

\DeclareRobustCommand{\IEEEauthorrefmark}[1]{\smash{\textsuperscript{\footnotesize #1}}}

\AddToShipoutPictureBG*{
  \AtPageUpperLeft{%
    \put(0,-40){\raisebox{15pt}{\makebox[\paperwidth]{\begin{minipage}{21cm}\centering
      \textcolor{gray}{This article has been accepted for publication in the proceedings of the \\
      27th International International Symposium on Power Electronics, Electrical Drives, Automation and Motion (SPEEDAM'24).\\ 
      } 
    \end{minipage}}}}%
  }
  \AtPageLowerLeft{%
    \raisebox{25pt}{\makebox[\paperwidth]{\begin{minipage}{21cm}\centering
      \textcolor{gray}{ \copyright 2024  Authors and IEEE. 
       This is the author’s version of the work. It is posted here for your personal use. Not for redistribution. The definitive Version of Record will
        be published in Proceedings of the 27th International International Symposium on Power Electronics, Electrical Drives, Automation and Motion (SPEEDAM'24)
      }
    \end{minipage}}}%
  }
}

\title{A Passive and Asynchronous Wake-up Receiver for Acoustic Underwater Communication}

\author{
\IEEEauthorblockN{
Lukas Schulthess\IEEEauthorrefmark{1}, 
Philipp Mayer\IEEEauthorrefmark{1}, 
Luca Benini\IEEEauthorrefmark{1}\IEEEauthorrefmark{2},Michele Magno\IEEEauthorrefmark{1}}
\IEEEauthorblockA{\IEEEauthorrefmark{1}Dept. of Information Technology and Electrical Engineering, ETH Z\"{u}rich, Switzerland} 
\IEEEauthorblockA{\IEEEauthorrefmark{2}Dept. of Electrical, Electronic and Information Engineering, University of Bologna, Italy}
}

\maketitle

\begin{abstract}
Establishing reliable data exchange in an underwater domain using energy and power-efficient communication methods is crucial and challenging. Radio frequencies are absorbed by the salty and mineral-rich water and optical signals are obstructed and scattered after short distances.
In contrast, acoustic communication benefits from low absorption and enables communication over long distances.
Underwater communication must match low power and energy requirements as underwater sensor systems must have a long battery lifetime and need to work reliably due to their deployment and maintenance cost.
For long-term deployments, the sensors' overall power consumption is determined by the power consumption during idle state.
It can be reduced by integrating asynchronous always-on wake-up circuits with nano-watt power consumption.
However, this approach does reduce but not eliminate idle power consumption, leaving a margin for improvement.


This paper presents a passive and asynchronous wake-up receiver for acoustic underwater communication enabling zero-power always-on listening. 
Zero-power listening is achieved by combining energy and information transmission using a low-power wake-up receiver that extracts energy out of the acoustic signal and eliminates radio frontend idle consumption. 
In-field evaluations demonstrate that the wake-up circuit requires only 
\textbf{$\mathbf{\qty{63}{\micro\watt}}$} to detect and compare an 8-bit UUID at a data rate of 200 bps up to a distance of \qty{5}{\meter} and that the needed energy can directly be extracted from the acoustic signal.
\end{abstract}

\vspace{10pt}

\section{Introduction}\label{sec:introduction}
\begin{figure}[!t]
    \centering
    \includegraphics[width=\columnwidth]{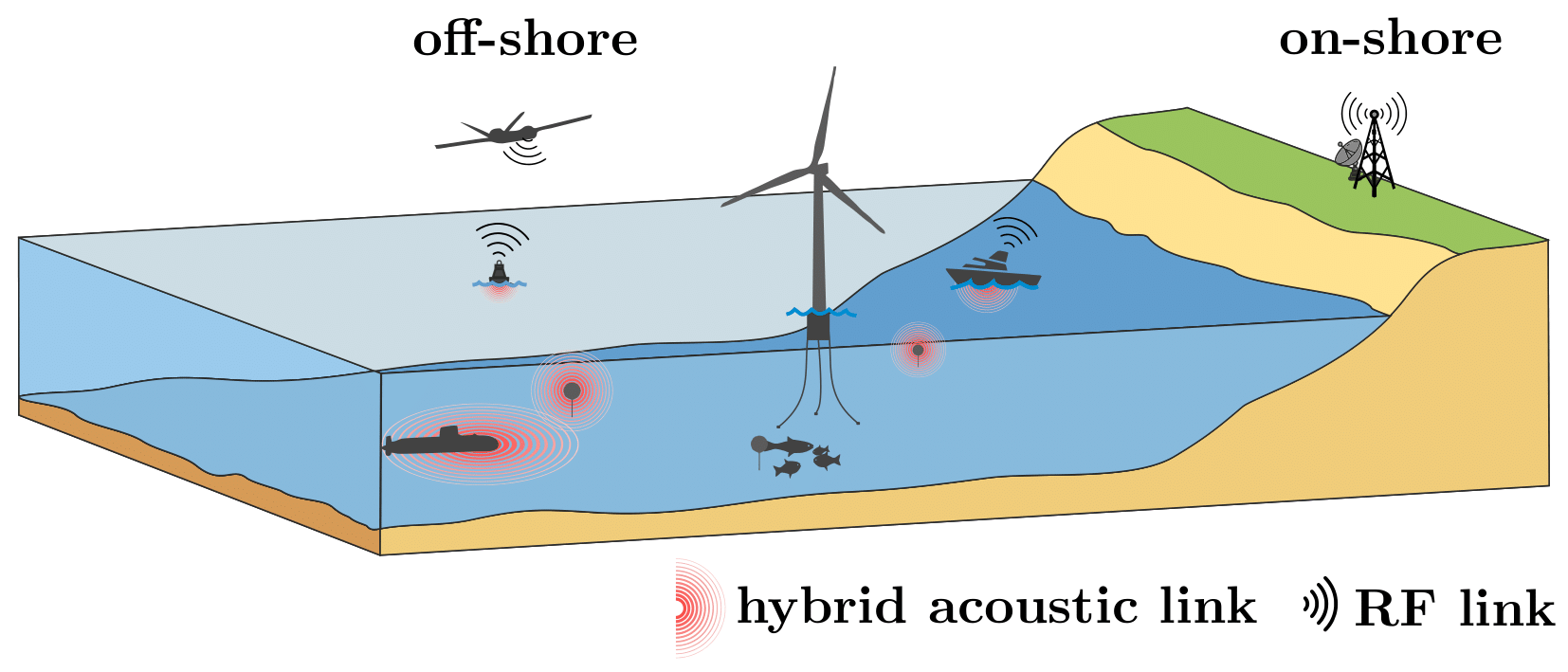}
    \vspace{-8mm}
    \caption{Application scenario overview with the proposed hybrid acoustic link as a backbone for underwater \ac{IoT}. The energy-neutral receiver allows a passive and asynchronous initialization of information transmission, thus increasing maritime sensor nodes' energy efficiency.}
    \vspace{-6mm}
    \label{fig:scenario}
\end{figure}

Oceans carry a rich variety of biological and non-biological resources. 
These resources are crucial for human life and therefore exists a great need for close and continuous monitoring \cite{schulthess_underwater_2023}. 
Traditionally, underwater sensors have been used to record data during their deployment. At the mission's conclusion, the sensor would be recovered for data analysis.
This approach is detrimental for time-critical monitoring tasks like surveillance or environmental monitoring as there is no possibility to provide feedback \cite{ferreira2023heterogeneous}.
Further, there is no possibility of detecting failures in the system during an ongoing recording, which can lead to an invalid monitoring mission \cite{schulthess_underwater_2023, berlian_2016}.
Therefore, there is a high demand for reliable underwater communication systems creating a \ac{UWN} for exchanging information.
Transferring data wirelessly between one or multiple sensor platforms \cite{yang2023research,cutinho_2022, demirors_2018} lays the foundation for various underwater activities in the fields of research \cite{cong2021underwater}, surveillance \cite{rumson_2021, xing_2021, terracciano_2020} and rescue \cite{cong2021underwater}.
Contrary to the traditional communication-less approaches, every node within the \ac{UWN} can collect and process environmental data instantaneously and, for example, locate and report unnatural pollution in coral reefs \cite{Coutinho2020} or provide early tsunami warnings \cite{inoue_2019}. 
Integrating \acp{AUV} into the \acp{UWN} extends its functionality, allows for precise and specific information collection, and enables numerous applications. The supervision of oil or gas pipelines on of-shore rigs and underwater construction monitoring \cite{terracciano_2020} can be executed completely autonomously, see \autoref{fig:scenario}.

The limited underwater usability of \ac{RF} \cite{Qureshi2016} and optical communication \cite{kaushal_2016} technologies combined with the increasing importance of industrial and environmental monitoring in the submarine and offshore environment implies significant innovation potential \cite{chaudhary2022underwater}.  
For long-range underwater communication, acoustic underwater channels in the range of hundred hertz up to several kilohertz are considered the gold standard \cite{stojanovic_2003} and have been implemented in acoustic modems \cite{wills_2006, sanchez_2012, jang_2019, renner_2020, hongbin_2022, wu_ic_nowbahari_2023}. 
Within this frequency range, acoustic signal propagation benefits from the water's low absorption rate, achieving communication distances over several kilometers at moderate data rates \cite{francois_1982_I, francois_1982_II}.
However, besides reliable communication, underwater sensor systems must have
a long battery lifetime and work reliably for years as their deployment
and maintenance are costly \cite{gentili2023comparison}.
To enable prolonged deployment, it is common practice to reduce active periods, thereby making the system's mission lifetime heavily dependent on its idle power consumption \cite{stewart2023performance,aoudia2016generic,piyare2018demand}. 
Ideally, the system should consume no power in the idle state but still be reactive to external communication requests \cite{gentili2023comparison}.
Periodic sensor activation, so-called "duty cycling", is not expedient as asynchronous communication events, i.e., passing drones, ships, or submarines are integral to most application scenarios \cite{wu_ic_nowbahari_2023}.
Integrating asynchronous wake-up circuits that consume power in the nano-watt range into the communication front-end can further improve the operational lifetime \cite{Magno2017}.
Although this strategy does decrease idle power consumption, it does not eliminate it.

This work presents a novel power-aware architecture for passive and asynchronous acoustic wake-up receivers. It overcomes the limitations of existing energy-efficient always-on underwater communication by combining a sub-system to harvest energy directly from acoustic messages with a self-powered, ultra-low power circuit for message reception and address decoding.
In particular, this article presents three novel contributions.
\begin{enumerate}
    \item Exploiting a piezoelectric transducer for simultaneously obtaining data and extracting power from the acoustic signal.
    \item Combining energy harvesting capabilities with a low-power wake-up receiver to achieve fully passive and asynchronous event reception with zero idle power consumption.
    \item Demonstrating the functionality of the novel system by extracting energy from the received message to power up the wake-up circuit and decoding an 8-bit UUID at a data rate of 200 bps up to a distance of \qty{5}{\meter}.
\end{enumerate}


\section{Related Work}\label{sec:relatedwork}
Whereas the challenges in establishing reliable communication over underwater acoustic channels have been well explored \cite{stewart2023performance, gentili2023comparison, yang2023research}, there is still potential in increasing the underwater sensor's lifetime using acoustic wake-up circuits \cite{erdem_2020}.
Recent trends in research focus on fully battery-less systems that are powered using acoustic energy harvesting and communicate using back-scattering \cite{us_backsctter_bhardway_2023, uw_backscatter_eid_2023, uw_backscatter_akbar_2023}. Although this approach can solve the limited power availability challenge, data can only be collected when a strong acoustic power source is present. Emitting loud and constant acoustic tones impacts the maritime environment and species \cite{acoustic_pollution_weilgart_2018, acoustic_pollution_stanley_2017}. 

Designing an \ac{UWSN} which minimizes power consumption in idle state and achieves a long battery-life time is important for systems that are deployed in harsh or remote locations where maintenance and access are limited.
Integrating energy harvesting techniques alongside wake-up functionalities \cite{pegatoquet2018wake} and ultra-low power designs has been shown to enhance the operational longevity of sensing platforms substantially. This synergistic approach, as highlighted in previous works \cite{mayer_2019, mayer_2018}, underscores the potential for significant efficiency improvements in such systems.
A comparison of both, academic and commercial state-of-the-art underwater acoustic wake-up receivers is presented in \autoref{tbl:uwsn_comparison}. 
It shows that most modern modems consume hundreds of microwatts of power in idle state \cite{renner_2020, hongbin_2022}. 
In fact, this power consumption is a factor of a hundred too large for the targeted maintenance-free operation, highlighting the need for passive and asynchronous wake-up receivers in acoustic underwater communication. 

\begin{table*}[!t]
\centering
\renewcommand{\arraystretch}{1.5}
\caption{Comparison of state-of-the-art underwater acoustic wake-up receivers}
     \begin{threeparttable}
     \begin{tabular}{p{2.4cm}P{1.6cm}P{1.6cm}P{1.8cm}P{1.6cm}P{1.7cm}P{1.9cm}P{2.0cm}}
     \toprule \\[-1.5em]
     & \makecell{\cite{wills_2006}\\2006} & \makecell{\cite{sanchez_2012}\\2012} & \makecell{\cite{jang_2019}\\2019} & \makecell{\cite{renner_2020}\\2020} & \makecell{\cite{hongbin_2022}\\2022} & \makecell{\cite{wu_ic_nowbahari_2023}\\2023 - IC} & 
     \makecell{\textbf{This Work}} \\
    \\[-1.5em]\midrule\\[-1.5em]
     Modulation & FSK & FSK & \makecell{Back-\\Scattering} & FSK & MFSK & OOK & 
     FSK \\
    \\[-1.5em]\\[-1.5em]
     \makecell[l]{$f_{carrier}\tnote{a}$\enspace(\si{\kilo\hertz})} & 18 & 85 &15 & 90 & 40 & 40 & 
     28 \\
     \\[-1.5em]\\[-1.5em]
     \makecell[l]{Data-Rate (kbps)} & N/A & 1 & \textbf{3} & 2.35 & 0.3 & 0.25 & 
     0.2 for UUID \\
     \\[-1.5em]\\[-1.5em]
     \makecell[l]{Detection Scheme} &  \makecell{Async.\\Wake-Up} & \makecell{Async.\\Wake-Up} & \makecell{Async.\\Wake-Up} & N/A & \makecell{Async.\\Wake-Up} & \makecell{Async.\\Wake-Up} & 
     \makecell{\textbf{Zero-Power}\\ \textbf{Wake-Up}} \\
     \\[-1.5em]\\[-1.5em]
      \makecell[l]{Battery-Less} & No & No & Yes & No & No & No & 
      Yes \\
     \\[-1.5em]\\[-1.5em]
     $P_{active}$\tnote{b} & N/A &  \SI{24}{\milli\watt} & \SI{500}{\micro\watt} & \SI{99}{\milli\watt} & \SI{600}{\milli\watt} & \textbf{\qty[text-series-to-math]{265}{\nano\watt}} & 
     \SI{63}{\micro\watt} \\
     \\[-1.5em]\\[-1.5em]
     $P_{idle}$\tnote{c} & \SI{500}{\micro\watt} & \SI{11}{\micro\watt} & \SI{125}{\micro\watt} & \SI{200}{\micro\watt} & \SI{2.4}{\milli\watt} &   \SI{61}{\nano\watt} & 
     \textbf{\qty[text-series-to-math]{0}{\micro\watt}} \\
     \\[-1.5em]\\[-1.5em]
     Tx-Stage &  Class D & Class B & Half-bridge & N/A & N/A & N/A & 
     Full-Bridge \\
     \\[-1.5em]\\[-1.5em]
     \makecell[l]{Wake-up Distance} & N/A & \SI{240}{\meter} & up to \SI{10}{\meter} & \SI{150}{\meter} & \SI{500}{\meter} & up to \SI{5}{\meter} & 
     \textbf{\makecell{up to \SI{5}{\meter}}} \\
     \\[-1.5em]\bottomrule\\[-1.5em]
     \end{tabular}
     \label{tbl:uwsn_comparison}
     \begin{tablenotes}
             \item[a] Used carrier frequency to establish acoustic communication. For FSK, the center frequency of all underwater acoustic channels has been reported.
             \item[b] $P_{active}$; Power consumed by the receiver when the system is active.
             \item[c] $P_{idle}$; Power consumption when the system is idle and waits for an acoustic wake-up signal.
     \end{tablenotes}  
     \vspace{-4mm}
 \end{threeparttable}   
 \end{table*}

\section{Communication Concept}
\begin{figure}[!t]
    \centering
    \begin{overpic}[width=\columnwidth]{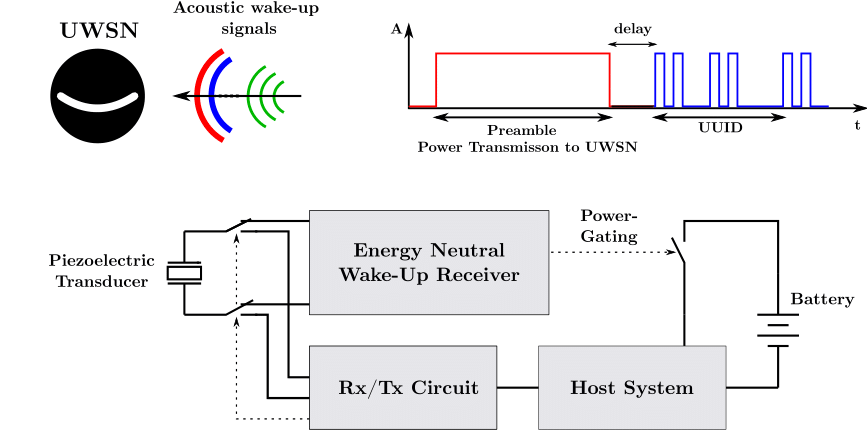}
        \put(1,50){(a)}
        \put(40,50){(b)}
        \put(1,23){(c)}
    \end{overpic}    \caption{(a) The communication is divided into two parts: A wake-up part (red and blue), and a part with the actual information (green). (b) Communication strategy used to power and wake individual sensor nodes. (c) A high-level overview of an underwater sensor node hosting an acoustic wake-up circuit connected to a single piezoelectric transducer.}
    \vspace{-4mm}
    \label{fig:topology}
\end{figure}

The proposed communication concept, as well as the system's high-level topology, is presented in \autoref{fig:topology}.
Dividing the actual communication into a wake-up part containing an \ac{UUID} to address a specific \ac{UWSN} (blue), and a longer part with the actual information (green), allows processing them separately (\autoref{fig:topology} (a)). 
Selecting different modulation schemes for wake-up and data brings benefits for power consumption and data rate.
A simple and power-efficient modulation scheme with low data rates can be used for the wake-up part. However, a more energy-hungry and throughput-efficient modulation can be exploited for information transfer once the system is awake \cite{schulthess_underwater_2023}.\\
In this work, the focus lies on designing a passive asynchronous wake-up receiver, that allows to power-gate the host system and thus efficaciously reduces the system's idle power consumption.
In idle state, the piezoelectric transducer is connected to the passive wake-up receiver and waits for a specific acoustic activation pattern. 
The wake-up signal is composed of a constant acoustic signal containing no data but carrying acoustic energy, followed by the sensor-specific \ac{UUID}.
The power-delivering preamble is formed out of a narrow-band acoustic tone to power up the battery-less wake-up receiver such that it can decode the subsequent \ac{OOK} modulated \ac{UUID}, \autoref{fig:topology} (b).
If the received \ac{UUID} complies with the sensor nodes \ac{UUID}, the host system is enabled, allowing the execution of advanced communication and processing strategies with increased data throughput, \autoref{fig:topology} (c). 

To implement this communication concept, the circuitry needs to be able to harvest energy from incoming acoustic signals, in this case from the dedicated preamble.
At the same time, the dynamic power consumption of the decoder sub-circuitry should be reduced as much as possible to lower the minimum amount of energy to be harvested for a successful \ac{UUID} decoding and thus extending the distance where passive wake-up is possible.
This can be achieved by using ultra-low power active components, passive components with high quality factors for the filter stages, and a high input impedance for the decoding part.
Further, the selected modulation strategy for data transmission directly influences the dynamic power consumption of the decoder. 
A simple modulation strategy for sending the \ac{UUID} reduced the demodulator complexity, as well as the associated power consumption, and thus is highly desirable when it comes to saving energy.
This work uses \ac{OOK} to implement a passive wake-up receiver. 
Contrary to \ac{FSK}, sending a \textit{binary zero} using \ac{OOK} costs no energy, leading to reduced energy needs for message transmissions. 
On the receiver's side, decoding an \ac{OOK}-modulated message only requires a small circuit demanding minimal power.

\section{Piezoelectric Transducer}
In underwater scenarios, the flow of the water is constantly changing in both speed and direction.
Axial rotation of the \ac{UWSN} can easily happen, even if the sensor is attached to an anchor or buoy. 
For this reason, a directed acoustic communication link is difficult to establish and has been considered unreliable. 
Thus, an omnidirectional receiver promises reliable performance for signal reception, independent of its axial rotation and relative position to the transmitter. \\
A piezoelectric transducer in the shape of a hemisphere fulfills this requirement and has been integrated. 
It has an outer diameter of \SI{80}{\milli\meter}, a wall thickness of \SI{3}{\milli\meter} and a total weight of \SI{211}{\gram}. 
Transmission and reception of acoustic signals are realized over the same piezoelectric hard \ac{PZT} transducer of navy type \RNum{3} with a resonant frequency of \SI{28}{\kilo\hertz}, allowing a reduction in device cost and higher integration.
During the idle state, the transducer is by design connected to the wake-up receiver and does not consume any power thanks to the mechanical reed relay switches, \autoref{fig:cirucit_blocks} (a), S\textsubscript{1}, S\textsubscript{2}. \\
To protect the silver-coated surface of the transducer against corrosion, which is especially important in salty waters, passive corrosion protection in the form of an environmentally compatible and highly insulating polyurethane coating using SOLRES01 from Scorpion Oceanics has been applied \cite{solres01}.
This specific polyurethane type has been selected to match the water's acoustic impedance as precisely as possible to maximize the sent and received acoustic energy
Using SOLRES01, a theoretical transmission coefficient of the polyurethane-water boundary $T\textsubscript{p-w}$ of 99.3 \% can be achieved.

\section{Hardware Architecture}
\begin{figure*}[!t]
    \centering
    \begin{overpic}[width=\textwidth]{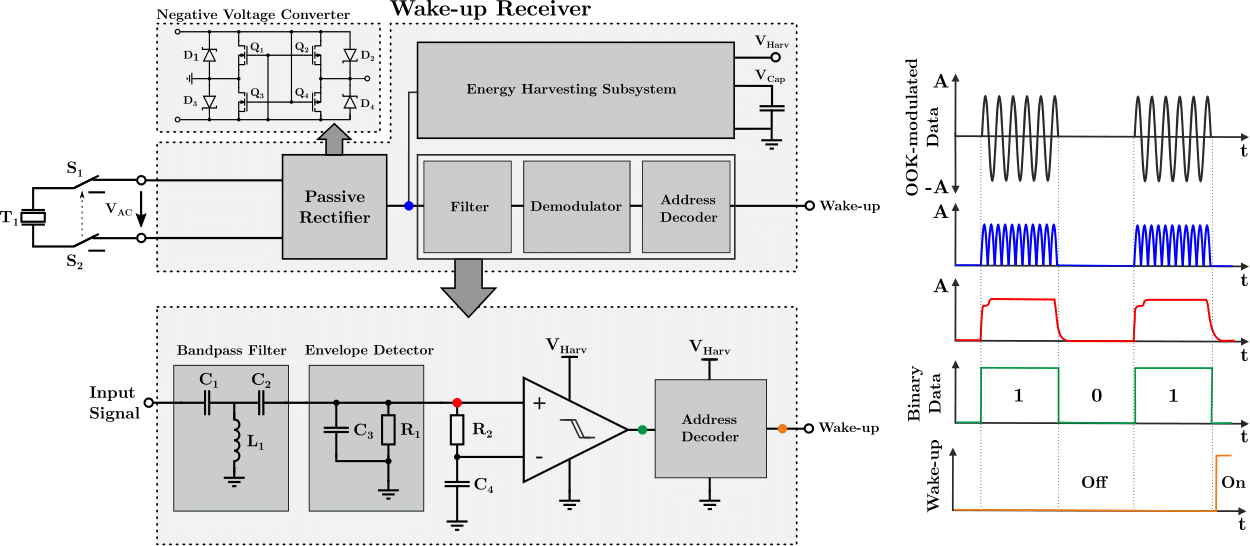}
        \put(2,42){(a)}
        \put(2,18){(b)}
        \put(72,42){(c)}
    \end{overpic}
    \caption{(a) High-level overview of the proposed acoustic wake-up circuit. (b) Detailed view of the \ac{OOK} demodulation circuit. (c) Involved signals in the demodulation process.}
    \label{fig:cirucit_blocks}
\end{figure*}

\autoref{fig:cirucit_blocks} (a) illustrates the passive wake-up receiver's high-level block diagram. It is composed of two sub-systems; the energy harvesting subsystem with intermittent energy storage and the \ac{OOK} demodulator for decoding the \ac{UUID}. 
Both are preceded by a passive rectifier to transform the AC signal from the piezoelectric transducer into a pulsating DC signal.

\textit{Passive Rectifier:} The piezoelectric transducer converts the modulated acoustic waves into an electric signal. 
To retrieve the underlying information and power as efficiently as possible, the acoustic signals are rectified using a \ac{NVCR} \cite{spu_mayer_2020}.
It combines a \ac{FWR} built out of Schottky diodes with a MOSFET-based \ac{NVC}.
For low input voltages below the transistors threshold voltage V\textsubscript{th}, only the Schottky diodes of type DSF01S30SL are active and allow small signal conversion. If the signal's amplitude exceeds the transistors V\textsubscript{th}, the transistors of the types CSD13383F4 and CSD23382F4 become conductive and the losses caused by the diode's forward voltage are reduced.

\textit{Energy Harvesting Subsystem:} 
To extract power from the rectified signal, an energy harvester is used for voltage conversion and intermediate energy buffering. 
Since it is directly connected after the \ac{NVCR}, energy can be harvested from a wide frequency range of received acoustic signals.
This allows accessing other signal sources for energy extraction, irrespective of their artificial or natural origin.
The energy harvesting subsystem is based on a BQ25570 energy harvester IC from Texas Instruments. 
To exit the harvester's energy-neutral idle state when the storage capacitor is completely depleted, a minimal input power of \SI{15}{\micro\watt} and a minimum input voltage of \SI{600}{\milli\volt} is needed. 
Those requirements define the minimal startup threshold for the energy-neutral wake-up receiver. 
Once the harvester is awake, the internal DC-DC boost charger allows harvesting energy from input voltages as low as \SI{100}{\milli\volt}. 
The for light loads optimized buck converter then generates a constant system voltage V\textsubscript{harv} of \SI{1.8}{\volt} to temporarily power up the active components of the preamble decoder.

\textit{Demodulator Subsystem:} 
To extract information from the \ac{OOK}-modulated signal, a power-efficient demodulating circuit has been implemented.
It consists of a passive band-pass filter, an envelope detector, a nano-power comparator, and an address decoder, \autoref{fig:cirucit_blocks} (b). 
The passive band-pass filter after the rectifier is designed to only let the \qty{28}{\kilo\hertz} carrier frequency of the \ac{OOK}-modulated \ac{UUID} pass, rejecting all the background noise from various maritime sources. 
The filter's T-matching network topology and component values are selected in such a way that the low-impedance signal source and the high-impedance of the decoder circuit are matched.
Next, an envelope detector smoothens out the high-frequency \ac{OOK} pulses to one single signal spike (ref. \autoref{fig:cirucit_blocks} (c), red). \\
A second low-pass filter right before the compactor's inputs lets the two inputs rise at different speeds, creating a voltage difference for incoming spikes. 
The nano-power comparator TLV3691 then translates this small analog voltage difference to defined digital output levels of the harvester's system voltage V\textsubscript{harv}, \autoref{fig:cirucit_blocks} (c), green. \\
Finally, the ultra-low power and lightweight \ac{MCU} PIC16LF15313 implements the address decoder. It compares the binary data stream generated by the comparator with its assigned \ac{UUID}. 
If they are identical, a \ac{GPIO} pin is set by the \ac{MCU}, generating a wake-up signal to deactivate the power-gating and enable further circuitry of the host system, \autoref{fig:cirucit_blocks} (c), orange.\\
To test the system, a hardware prototype has been designed. 
It consists of two custom \acp{PCB} stacked on top of each other, the piezoelectric transducer with a protective coating, and a top cover as depicted in \autoref{fig:test_setup} (a).
The lower \ac{PCB} hosts the fully asynchronous passive wake-up system, as well as a tx-stage and two reed relays to decouple the latter from the power-sensitive wake-up circuit. 
The wake-up signal is accessible over a pin connector and allows for a modular system, extending any host system with underwater communication and passive wake-up functionalities. 
The upper \ac{PCB} implements the custom host system and includes the power gating hardware.


\section{Experimental Results}\label{sec:results}
\begin{figure}[!ht]
    \centering
    \begin{overpic}[width=\columnwidth]{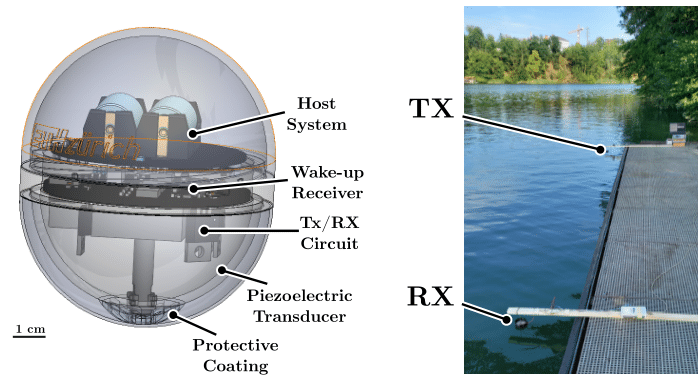}
        \put(2,51){(a)}
        \put(58,51){(b)}
    \end{overpic}
    \vspace{-6mm}
    \caption{(a) 3D rendering of the designed \ac{UWSN} which integrates the proposed wake-up receiver, the piezoelectric hemisphere for signal reception and energy extraction, and a more powerful host system. (b) Illustration of the test setup at the river site. 
    }
   \vspace{-5mm}
    \label{fig:test_setup}
\end{figure}

To evaluate the functionality of the proposed passive and asynchronous wake-up receiver, experimental results in a river with an approximate depth of \qty{5}{\meter} have been conducted. The test platform reaches \qty{2.6}{\meter} into the river and rests on two concrete posts (\autoref{fig:test_setup} (b)).
The acoustic signals have been generated by a piezoelectric hemisphere and a Tx circuit using \qty{200}{\volt} generated by an EA-EL 900 DT programmable DC source.
On the receiver side, an NI-USB 6216 isolated data acquisition device has been used for measurement.

\begin{figure}[!hb]
    \vspace{-2mm}
    \begin{overpic}[width=0.98\columnwidth, right]{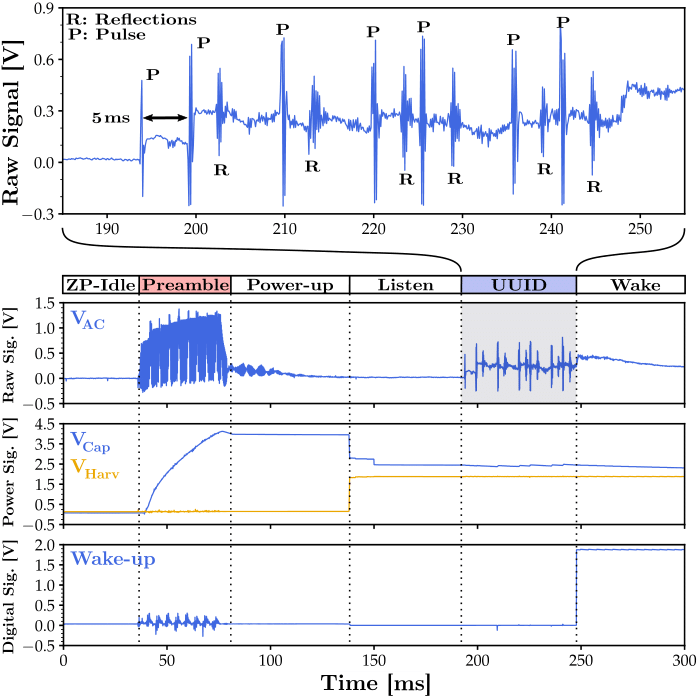}
        \put(1,95){(a)}
        \put(1,58){(b)}
    \end{overpic}
    \vspace{-7mm}
    \caption{Demonstration of the passive wake-up receiver. The preceding preamble provides energy to power up the system and decode the UUID. The wake-up signal is activated if the received UUID matches the wake-up receiver.}
    \vspace{-3mm}
    \label{fig:results}
\end{figure}

\autoref{fig:results} (b) demonstrates a complete wake-up cycle.
Initially, the wake-up receiver is in its passive idle state, and the capacitor for intermittent energy storage $V\textsubscript{Cap}$ depleted. 
To exit the idle state, an acoustic preamble of constant tone at \qty{28}{\kilo\hertz} is initially sent from the transmitter to wake-up the receiver.
When the cold-start conditions of the energy harvester have been fulfilled, the energy harvesting subsystem is activated.
It extracts energy from the acoustic preamble and stores it in the capacitance $V\textsubscript{Cap}$.
A preamble length of \qty{50}{\milli\second} contains enough energy to charge the auxiliary capacitance of \qty{100}{\micro\farad} to a maximum voltage of \qty{4.12}{\volt} and corresponds to a harvested energy of \qty{849}{\micro\joule} over a short distance of \qty{1}{\meter}. \\
By adopting the preamble duration, more energy can be transmitted to extend the wake-up distance or reduced to save transmission power. Experimental results report a stable wake-up functionality of up to \qty{5}{\meter} with a preamble length of \qty{400}{\milli\second}.
Once the system is powered up, it actively listens for the subsequent \ac{UUID} and consumes \qty{10.7}{\micro\watt}. 
The \ac{UUID} structure consists of 10 bits with the first two bits set to one.  
The wake-up system measures the delay between those bits and uses it as a reference delay to define the sampling time of the next bit. 

With this, unknown channel characteristics are considered and the system is not restricted to only one data rate for transmitting the \ac{UUID}.
In shallow water scenarios, hard surfaces like the bank or the bed of a river reflect acoustic signals.
Depending on the energy content and temporal appearance of these reflections, the original message can be distorted.
\autoref{fig:results} (a) indicates acoustic ones of the received \ac{UUID} as \textit{P} and their corresponding reflections as \textit{R}. The delay between a pulse and its reflection from the riverbank is \qty{3.1}{\milli\second}. 
Together with the underwater sound speed of \qty{1630}{\meter\per\second}, the reflective path is calculated to be \qty{5.05}{\meter} longer than the line-of-sight.
However, due to constant sampling, reflections only affect the signal when they fall into the sampling time.
The first critical distance from the transmitter to a receiver over a reflective surface is thus defined as the sampling period multiplied by the underwater sound speed.
For our system sending \ac{UUID} at 200 bps, this distance lies at \qty{8.15}{\meter}.
Reflections are mainly an issue for shallow-water scenarios and need to be considered during evaluation as most test setups fall into this category.

\section{Conclusion}\label{sec:conclusion}
This work proposed a passive always-on receiver architecture for asynchronous underwater communication. The design eliminates the constant power consumption in idle state by combining energy harvesting from the received acoustic message with a self- and low-power Rx-Tx circuitry for address decoding.
The proposed hybrid approach allows harvesting sufficient energy directly from the receiving signal to power a wake-up receiver stage, thus allowing a truly energy-neutral idle operation of the UWSN.
The power-efficient wake-up circuit design only requires \SI{63}{\micro\watt} in active mode to detect and compare an 8-bit UUID at a data rate of 200 bps.\\
To the best of our knowledge, all the previous proposed solutions either suffer from high energy consumption in idle mode or require constant acoustic power transfer. Our approach completely eliminates the idle currents of the communication interface and thus significantly prolongs the mission lifetime of underwater sensor nodes, the most critical feature for such devices. Our work allows for fully passive wake-ups and asynchronous communication while minimizing acoustic exposure to the maritime environment and improving the latency of the communication avoiding duty cycling.
Furthermore, the circuit has been built out of low-cost off-the-shelf components allowing easy integration and scaling.

\FloatBarrier


\begin{acronym}
    \acro{RF}{radio frequency}
    \acro{IoT}{internet of things}
    \acro{IoUT}{internet of underwater things}
    \acro{UWN}{underwater wireless network}
    \acro{UWSN}{underwater wireless sensor node}
    \acro{UAC}{underwater acoustic channel}
    \acro{FSK}{frequency shift keying}
    \acro{OOK}{on-off keying}
    \acro{ASK}{amplitude shift keying}
    \acro{UUID}{universal unique identifier}
    \acro{PZT}{lead zirconium titanate}
    \acro{AC}{alternating current}
    \acro{NVC}{negative Voltage converter}
    \acro{NVCR}{negative voltage converter rectifier}
    \acro{FWR}{full-Wave rectifier}
    \acro{MCU}{microcontroller}
    \acro{GPIO}{general purpose input/output}
    \acro{PCB}{printed circuit board}
    \acro{AUV}{autonomous underwater vehicle}

\end{acronym}
\bibliographystyle{IEEEtranDOI} 
\bibliography{main}

\begin{thebibliography}{10}
\providecommand{\url}[1]{#1}
\csname url@samestyle\endcsname
\providecommand{\newblock}{\relax}
\providecommand{\bibinfo}[2]{#2}
\providecommand{\BIBentrySTDinterwordspacing}{\spaceskip=0pt\relax}
\providecommand{\BIBentryALTinterwordstretchfactor}{4}
\providecommand{\BIBentryALTinterwordspacing}{\spaceskip=\fontdimen2\font plus
\BIBentryALTinterwordstretchfactor\fontdimen3\font minus \fontdimen4\font\relax}
\providecommand{\BIBforeignlanguage}[2]{{%
\expandafter\ifx\csname l@#1\endcsname\relax
\typeout{** WARNING: IEEEtran.bst: No hyphenation pattern has been}%
\typeout{** loaded for the language `#1'. Using the pattern for}%
\typeout{** the default language instead.}%
\else
\language=\csname l@#1\endcsname
\fi
#2}}
\providecommand{\BIBdecl}{\relax}
\BIBdecl

\bibitem{schulthess_underwater_2023}
L.~Schulthess, P.~Mayer, and M.~Magno, ``Poster abstract: Battery-free and passive wake-up receiver for underwater communication,'' in \emph{Proceedings of the 8th ACM/IEEE Conference on Internet of Things Design and Implementation (IoTDI '23)}, 2023, p. 479–480, doi: 10.1145/3576842.3589178.

\bibitem{ferreira2023heterogeneous}
F.~Ferreira \emph{et~al.}, ``Heterogeneous marine robotic system for environmental monitoring missions,'' in \emph{2023 IEEE Underwater Technology (UT)}, 2023, pp. 1--5, doi: 10.1109/UT49729.2023.10103383.

\bibitem{berlian_2016}
M.~H. Berlian \emph{et~al.}, ``{Design and implementation of smart environment monitoring and analytics in real-time system framework based on internet of underwater things and big data},'' in \emph{2016 International Electronics Symposium (IES)}, 2016, pp. 403--408, doi: 10.1109/ELECSYM.2016.7861040.

\bibitem{yang2023research}
X.~Yang, Y.~Zhou, R.~Wang, and F.~Tong, ``Research and implementation on a real-time osdm modem for underwater acoustic communications,'' \emph{IEEE Sensors Journal}, vol.~23, no.~16, pp. 18\,434--18\,448, 2023, doi: 10.1109/JSEN.2023.3291082.

\bibitem{cutinho_2022}
R.~W.~L. Coutinho, A.~Boukerche, L.~F.~M. Vieira, and A.~A.~F. Loureiro, ``Underwater sensor networks for smart disaster management,'' \emph{IEEE Consumer Electronics Magazine}, vol.~9, no.~2, pp. 107--114, 2020, doi: 10.1109/MCE.2019.2953686.

\bibitem{demirors_2018}
E.~Demirors \emph{et~al.}, ``{The SEANet project: Toward a programmable internet of underwater things},'' in \emph{2018 4th Underwater Communications and Networking Conference, UComms 2018}, oct 2018, doi: 10.1109/UComms.2018.8493207.

\bibitem{cong2021underwater}
Y.~Cong, C.~Gu, T.~Zhang, and Y.~Gao, ``Underwater robot sensing technology: A survey,'' \emph{Fundamental Research}, vol.~1, no.~3, pp. 337--345, 2021, doi: 10.1016/j.fmre.2021.03.002.

\bibitem{rumson_2021}
\BIBentryALTinterwordspacing
A.~G. Rumson, ``The application of fully unmanned robotic systems for inspection of subsea pipelines,'' \emph{Ocean Engineering}, vol. 235, p. 109214, 2021, doi: 10.1016/j.oceaneng.2021.109214.
\BIBentrySTDinterwordspacing

\bibitem{xing_2021}
H.~Xing \emph{et~al.}, ``{A Multi-Sensor Fusion Self-Localization System of a Miniature Underwater Robot in Structured and GPS-denied Environments},'' \emph{IEEE Sensors Journal}, pp. 1--1, oct 2021, doi: 10.1109/jsen.2021.3120663.

\bibitem{terracciano_2020}
D.~S. Terracciano, L.~Bazzarello, A.~Caiti, R.~Costanzi, and V.~Manzari, ``Marine robots for underwater surveillance,'' \emph{Current Robotics Reports}, vol.~1, 12 2020, doi: 10.1007/s43154-020-00028-z.

\bibitem{Coutinho2020}
R.~W. Coutinho, A.~Boukerche, L.~F. Vieira, and A.~A. Loureiro, ``{Underwater Sensor Networks for Smart Disaster Management},'' \emph{IEEE Consumer Electronics Magazine}, vol.~9, no.~2, pp. 107--114, mar 2020, doi: 10.1109/MCE.2019.2953686.

\bibitem{inoue_2019}
\BIBentryALTinterwordspacing
M.~Inoue, Y.~Tanioka, and Y.~Yamanaka, ``Method for near-real time estimation of tsunami sources using ocean bottom pressure sensor network (s-net),'' \emph{Geosciences}, vol.~9, no.~7, 2019, doi: 10.3390/geosciences9070310.
\BIBentrySTDinterwordspacing

\bibitem{Qureshi2016}
U.~M. Qureshi \emph{et~al.}, ``Rf path and absorption loss estimation for underwater wireless sensor networks in different water environments,'' \emph{Sensors}, vol.~16, no.~6, 2016, doi: 10.3390/s16060890.

\bibitem{kaushal_2016}
H.~Kaushal and G.~Kaddoum, ``Underwater optical wireless communication,'' \emph{IEEE Access}, vol.~4, pp. 1518--1547, 2016, doi: 10.1109/ACCESS.2016.2552538.

\bibitem{chaudhary2022underwater}
M.~Chaudhary, N.~Goyal, A.~Benslimane, L.~K. Awasthi, A.~Alwadain, and A.~Singh, ``Underwater wireless sensor networks: Enabling technologies for node deployment and data collection challenges,'' \emph{IEEE Internet of Things Journal}, vol.~10, no.~4, pp. 3500--3524, 2023, doi: 10.1109/JIOT.2022.3218766.

\bibitem{stojanovic_2003}
M.~Stojanovic, \emph{Acoustic (Underwater) Communications}.\hskip 1em plus 0.5em minus 0.4em\relax John Wiley \& Sons, Ltd, 2003, ISBN 9780471219286.

\bibitem{wills_2006}
J.~Wills, W.~Ye, and J.~Heidemann, ``{Low-power acoustic modem for dense underwater sensor networks},'' in \emph{WUWNet 2006 - Proceedings of the First ACM International Workshop on Underwater Networks}, vol. 2006, 2006, pp. 79--85, doi: 10.1145/1161039.1161055.

\bibitem{sanchez_2012}
A.~S{\'{a}}nchez, S.~Blanc, P.~Yuste, A.~Perles, and J.~Serrano~Mart{\'{i}}n, ``{An Ultra-Low Power and Flexible Acoustic Modem Design to Develop Energy-Efficient Underwater Sensor Networks},'' \emph{Sensors (Basel, Switzerland)}, vol.~12, pp. 6837--6856, 2012, doi: 10.3390/s120606837.

\bibitem{jang_2019}
\BIBentryALTinterwordspacing
J.~Jang and F.~Adib, ``Underwater backscatter networking,'' in \emph{Proceedings of the ACM Special Interest Group on Data Communication}, ser. SIGCOMM '19, New York, NY, USA, 2019, p. 187–199, doi: 10.1145/3341302.3342091.
\BIBentrySTDinterwordspacing

\bibitem{renner_2020}
\BIBentryALTinterwordspacing
B.-C. Renner, J.~Heitmann, and F.~Steinmetz, ``Ahoi: Inexpensive, low-power communication and localization for underwater sensor networks and uauvs,'' \emph{ACM Trans. Sen. Netw.}, vol.~16, no.~2, jan 2020, doi: 10.1145/3376921.
\BIBentrySTDinterwordspacing

\bibitem{hongbin_2022}
H.~Chen, Y.~Zhu, W.~Zhang, K.~Wu, and F.~Yuan, ``Underwater acoustic micromodem for underwater internet of things,'' \emph{Wireless Communications and Mobile Computing}, vol. 2022, pp. 1--20, 09 2022, doi: 10.1155/2022/9148756.

\bibitem{wu_ic_nowbahari_2023}
A.~Nowbahari, L.~Marchetti, and M.~Azadmehr, ``Low power wake-up receivers for underwater acoustic wireless sensor networks,'' \emph{IEEE Transactions on Green Communications and Networking}, vol.~7, no.~4, pp. 1635--1647, 2023, doi: 10.1109/TGCN.2023.3279627.

\bibitem{francois_1982_I}
R.~E. Francois and G.~R. Garrison, ``Sound absorption based on ocean measurements: Part i: Pure water and magnesium sulfate contributions,'' \emph{Journal of the Acoustical Society of America}, vol.~72, pp. 896--907, 1982.

\bibitem{francois_1982_II}
------, ``Sound absorption based on ocean measurements: Part ii: Boric acid contribution and equation for total absorption,'' \emph{Journal of the Acoustical Society of America}, vol.~72, pp. 1879--1890, 1982.

\bibitem{gentili2023comparison}
A.~Gentili \emph{et~al.}, ``Comparison of passive acoustic monitoring sensors for direction of arrival estimation of underwater acoustic sources,'' in \emph{2023 IEEE International Workshop on Metrology for the Sea; Learning to Measure Sea Health Parameters (MetroSea)}, 2023, pp. 534--539, doi: 10.1109/MetroSea58055.2023.10317252.

\bibitem{stewart2023performance}
C.~Stewart, N.~Fough, N.~Erdogan, and R.~Prabhu, ``Performance and energy modelling for a low energy acoustic network for the underwater internet of things,'' in \emph{2023 IEEE International Workshop on Metrology for the Sea; Learning to Measure Sea Health Parameters (MetroSea)}, 2023, pp. 110--115, doi: 10.1109/MetroSea58055.2023.10317266.

\bibitem{aoudia2016generic}
F.~A. Aoudia, M.~Gautier, M.~Magno, O.~Berder, and L.~Benini, ``A generic framework for modeling mac protocols in wireless sensor networks,'' \emph{IEEE/ACM Transactions on Networking}, vol.~25, no.~3, pp. 1489--1500, 2016.

\bibitem{piyare2018demand}
R.~Piyare, A.~L. Murphy, M.~Magno, and L.~Benini, ``On-demand tdma for energy efficient data collection with lora and wake-up receiver,'' in \emph{2018 14th International Conference on Wireless and Mobile Computing, Networking and Communications (WiMob)}.\hskip 1em plus 0.5em minus 0.4em\relax IEEE, 2018, pp. 1--4.

\bibitem{Magno2017}
M.~Magno and D.~Boyle, ``{Wearable Energy Harvesting: From body to battery},'' in \emph{Proceedings - 2017 12th IEEE International Conference on Design and Technology of Integrated Systems in Nanoscale Era, DTIS 2017}, 2017, doi: 10.1109/DTIS.2017.7930169.

\bibitem{erdem_2020}
\BIBentryALTinterwordspacing
H.~E. Erdem and V.~C. Gungor, ``Analyzing lifetime of energy harvesting underwater wireless sensor nodes,'' \emph{International Journal of Communication Systems}, vol.~33, no.~3, p. e4214, 2020, doi: https://doi.org/10.1002/dac.4214, E4214 IJCS-19-0517.R1.
\BIBentrySTDinterwordspacing

\bibitem{us_backsctter_bhardway_2023}
A.~Bhardwaj, A.~Allam, A.~Erturk, and K.~G. Sabra, ``Ultrasound-powered wireless underwater acoustic identification tags for backscatter communication,'' \emph{IEEE Transactions on Ultrasonics, Ferroelectrics, and Frequency Control}, vol.~71, pp. 304--313, 2024, doi: 10.1109/TUFFC.2023.3344638.

\bibitem{uw_backscatter_eid_2023}
\BIBentryALTinterwordspacing
A.~Eid, J.~Rademacher, W.~Akbar, P.~Wang, A.~Allam, and F.~Adib, ``Enabling long-range underwater backscatter via van atta acoustic networks,'' in \emph{Proceedings of the ACM SIGCOMM 2023 Conference}, ser. ACM SIGCOMM '23, New York, NY, USA, 2023, p. 1–19, doi: 10.1145/3603269.3604814.
\BIBentrySTDinterwordspacing

\bibitem{uw_backscatter_akbar_2023}
W.~Akbar, A.~Allam, and F.~Adib, \emph{The Underwater Backscatter Channel: Theory, Link Budget, and Experimental Validation}.\hskip 1em plus 0.5em minus 0.4em\relax New York, NY, USA: Association for Computing Machinery, 2023, ISBN 9781450399906.

\bibitem{acoustic_pollution_weilgart_2018}
L.~Weilgart, ``The impact of ocean noise pollution on fish and invertebrates,'' \emph{OceanCare}, p.~36, 2018.

\bibitem{acoustic_pollution_stanley_2017}
J.~Stanley, S.~Van~Parijs, and L.~Hatch, ``Underwater sound from vessel traffic reduces the effective communication range in atlantic cod and haddock,'' \emph{Scientific Reports}, vol.~7, 11 2017, doi: 10.1038/s41598-017-14743-9.

\bibitem{pegatoquet2018wake}
A.~Pegatoquet, T.~N. Le, and M.~Magno, ``A wake-up radio-based mac protocol for autonomous wireless sensor networks,'' \emph{IEEE/ACM Transactions on Networking}, vol.~27, no.~1, pp. 56--70, 2018.

\bibitem{mayer_2019}
P.~Mayer, M.~Magno, and L.~Benini, ``{Self-Sustaining Acoustic Sensor with Programmable Pattern Recognition for Underwater Monitoring},'' \emph{IEEE Transactions on Instrumentation and Measurement}, vol.~68, no.~7, pp. 2346--2355, jul 2019, doi: 10.1109/TIM.2018.2890187.

\bibitem{mayer_2018}
------, ``{Combining microbial fuel cell and ultra-low power event-driven audio detector for zero-power sensing in underwater monitoring},'' in \emph{2018 IEEE Sensors Applications Symposium (SAS)}, 2018, pp. 1--6, doi: 10.1109/SAS.2018.8336772.

\bibitem{solres01}
{Scorpion Oceanics}, ``Polyurethane mould materials,'' \url{https://www.scorpionoceanics.co.uk/resources/}, accessed: 2024-03-014.

\bibitem{spu_mayer_2020}
P.~Mayer, M.~Magno, and L.~Benini, ``Smart power unit—mw-to-nw power management and control for self-sustainable iot devices,'' \emph{IEEE Transactions on Power Electronics}, vol.~36, no.~5, pp. 5700--5710, 2021, doi: 10.1109/TPEL.2020.3031697.

\end{thebibliography}

\end{document}